\documentclass[aps,prb,twocolumn,superscriptaddress,showpacs]{revtex4-1}
\usepackage[pdftex]{graphicx} %%%MIGHT HAVE TO BE CHANGED TO ACCEPT *.eps FILES!!!!!!!
\usepackage{amsmath} %for align-environment
\usepackage{amssymb} %for \triangledown and \bigstar
\usepackage{mathtools}
\usepackage{color} %for colorful symbols
\usepackage{textcomp} %for \textpm in appendix
\usepackage{url}

\begin{document}

%\preprint{}

%Title of paper %first letter capitalized, others lower-case
\title{Scattering cross sections of liquid deuterium for ultracold neutrons: Experimental results and a calculation model}



\author{Stefan D\"{o}ge}
\email[]{stefan.doege@tum.de}
%\thanks{}
\affiliation{Physik-Department, Technische Universit\"{a}t M\"{u}nchen, D-85748 Garching, Germany}
\affiliation{Institut Laue--Langevin, 71 avenue des Martyrs, F-38042 Grenoble Cedex 9, France}
\affiliation{\'{E}cole Doctorale de Physique, Universit\'{e} Grenoble Alpes, F-38402 Saint Martin d'H\`{e}res, France}

\author{Christian Herold}
\affiliation{RWTH Aachen, NET, D-52062 Aachen, Germany}

\author{Stefan M\"{u}ller}
\affiliation{Physik-Department, Technische Universit\"{a}t M\"{u}nchen, D-85748 Garching, Germany}

\author{Christoph Morkel}
\affiliation{Physik-Department, Technische Universit\"{a}t M\"{u}nchen, D-85748 Garching, Germany}

\author{Erwin Gutsmiedl}
\affiliation{Physik-Department, Technische Universit\"{a}t M\"{u}nchen, D-85748 Garching, Germany}

\author{Peter Geltenbort}
\affiliation{Institut Laue--Langevin, 71 avenue des Martyrs, F-38042 Grenoble Cedex 9, France}

\author{Thorsten Lauer}
\affiliation{Forschungsneutronenquelle Heinz Maier-Leibnitz, Technische Universit\"{a}t M\"{u}nchen, D-85748 Garching, Germany}

\author{Peter Fierlinger}
\affiliation{Physik-Department, Technische Universit\"{a}t M\"{u}nchen, D-85748 Garching, Germany}

\author{Winfried Petry}
\affiliation{Forschungsneutronenquelle Heinz Maier-Leibnitz, Technische Universit\"{a}t M\"{u}nchen, D-85748 Garching, Germany}

\author{Peter B\"{o}ni}
\affiliation{Physik-Department, Technische Universit\"{a}t M\"{u}nchen, D-85748 Garching, Germany}


\date{Published as Phys. Rev. B 91, 214309 on June 24, 2015} %\date{\today}

\begin{abstract} %max. 600 characters incl. spacing
We present scattering cross sections $\sigma_{\text{scatt}}$ of ultracold neutrons (UCN) in liquid deuterium at $T = 20.6\,\text{K}$, as recently measured by means of a transmission experiment. The indispensable thorough raw data treatment procedure is explained. A calculation model for coherent and incoherent scattering in liquid deuterium in the hydrodynamic limit based on appropriate physical concepts is provided and shown to fit the data well. The applicability of the incoherent approximation for UCN scattering in liquid deuterium was tested and found to deliver acceptable results.
\end{abstract}

%http://www.aip.org/publishing/pacs/pacs-reg20#28
\pacs{28.20.Cz, 28.41.Pa, 61.25.Em, 61.05.fd} %28.20.Cz (Neutron scattering), 28.41.Pa (Moderators), 24.10.Nz (Hydrodynamic models), 25.40.Dn (Elastic neutron scattering), 25.40.Fq (Inelastic neutron scattering), 61.25.Em (Molecular liquids), 61.05.fd (Theories of neutron diffraction and scattering)
%\keywords{}

\maketitle

%\section{Introduction}
%\subsection{Intro Subsection}
%\subsubsection{Intro Subsubsection}

\section{Introduction}
Liquid deuterium is used as a moderator for thermal neutrons in cold neutron sources at neutron research facilities (e.g., For\-schungs\-neu\-tro\-nen\-quelle Heinz Maier-Leibnitz (FRM II) \citep{tum-frm2:2014} in Garching, Germany, and Institut Laue--Langevin (ILL) \citep{cea:2012} in Grenoble, France). They are operated at around 25\,K and deliver a cold neutron spectrum with a mean neutron kinetic energy of 2.2\,meV. Their output spectrum can be more accurately calculated when using an extension of the scattering data down to the ultracold neutron (UCN) range. Scattering cross section measurements in liquid deuterium have been performed for thermal and cold neutrons \citep{seiffert:1970}, as well as for very cold (VCN) and ultracold (UCNs) neutrons \citep{atchison:2005-liq}. However, there appears to be no comprehensive raw data treatment for transmission experiments, especially in the UCN range, so far described, and no theoretical model yet to explain the scattering cross section behavior. The research presented here proposes both a raw data treatment and a theoretical explanation of the scattering cross sections \citep{doege:2014}.

UCNs are neutrons slow enough to be reflected from suitable materials under any angle of incidence. This allows for their confinement in material bottles, which significantly increases interaction and observation times. Typically, UCNs are defined to have a kinetic energy of $\sim$\,350\,neV or less. In our research, neutrons with an energy lower than 1000\,neV are regarded as ultracold, for this is the maximum energy of the neutron spectrum supplied by the UCN source (``turbine") at ILL \citep{steyerl:1975, steyerl:1986}. UCNs are used predominantly in the study of fundamental physics principles, such as the free-neutron lifetime \citep{yue:2013}, the search for a possible nonzero neutron electric dipole moment (nEDM) \citep{baker:2011}, and the validity of Newton's law of gravity on the micrometer scale. The former two experiments are valuable tools to test the standard model of particle physics and its extensions or even to discover physics beyond the standard model. Another research project benefiting from the use of UCNs is the measurement of the neutron $\beta$-decay asymmetry \citep{mendenhall:2013}, which helps us to understand the spin and flavor structure of the nucleon. At present, statistics are the main limiting factor of these experiments. Significantly higher UCN fluxes will allow a considerable increase of these experiments' accuracy. To date, ILL's turbine is the most intense UCN source, while solid deuterium converters are promising to increase the maximum possible flux in the near future \citep{bodek:2008, lauss:2014}. For further references and the history of UCN science, see the books of Ignatovich \citep{ignatovich:1990} and Golub \citep{golub:1991}.

\section{Experimental Setup}
Due to the nuclear spin of the deuteron ($I$=1) the homonuclear diatomic deuterium molecule exists in two symmetry configurations (also called \emph{species}) -- \textit{ortho} (with an even rotational quantum number $J$) and \textit{para} (odd $J$) \citep{silvera:1980}. In order to suppress neutron up-scattering from the rotational relaxation $J$=1$\rightarrow$0, the \textit{ortho}-deuterium concentration was maximized (i.e., to $c_{\text{o}}\, \simeq\, 0.98$) in previous experiments examining the viability of deuterium and deuterated substances as UCN converters \citep{atchison:2005-liq, atchison:2005-sol, atchison:2011}. In our experiment we used enriched \textit{ortho}-deuterium with $c_{\text{o}}\, =\, 0.80$, which was prepared in an Oxisorb$^{\text{\textregistered}}$ converter (CrO$_{\text{3}}$-based) in residence mode at about 20\,K, similar to the one described by Bodek et al. \citep{bodek:2004} In the converter, chromium trioxide CrO$_{\text{3}}$ bonded to silica gel provides paramagnetic centers and thus magnetic field gradients, which catalyze the otherwise slow conversion towards the low-temperature equilibrium.

The experiment described here is a pure transmission experiment; the direction of the momentum transfer is not being determined. The experiment yields the total cross section $\sigma_\text{tot}$ of liquid deuterium through the following equations
\begin{equation}
\frac{I_\text{filled}}{I_\text{vacuum}} = e^{-\Sigma_\text{tot}d} = e^{-\sigma_\text{tot} N_\text{V} d}
\end{equation}

\begin{equation}
\label{eq:transmission}
\sigma_\text{tot} = \frac{1}{N_\text{V} d} \ln \left( \frac{I_\text{vacuum}}{I_\text{filled}} \right),
\end{equation}
where $I$ is the UCN count rate at the detector (with the subscript \emph{vacuum} for vacuum in the sample cell and \emph{filled} with liquid deuterium), $d$ is the sample cell thickness, and $N_\text{V}$ the number density of deuterium molecules. The condition $\Sigma_\text{tot} d < 1$ must be met in order to reduce multiple scattering and not to distort the \emph{single} scattering cross section, which is measured in this experiment.

\begin{figure}[!h]
	\includegraphics[width=1.00\columnwidth]{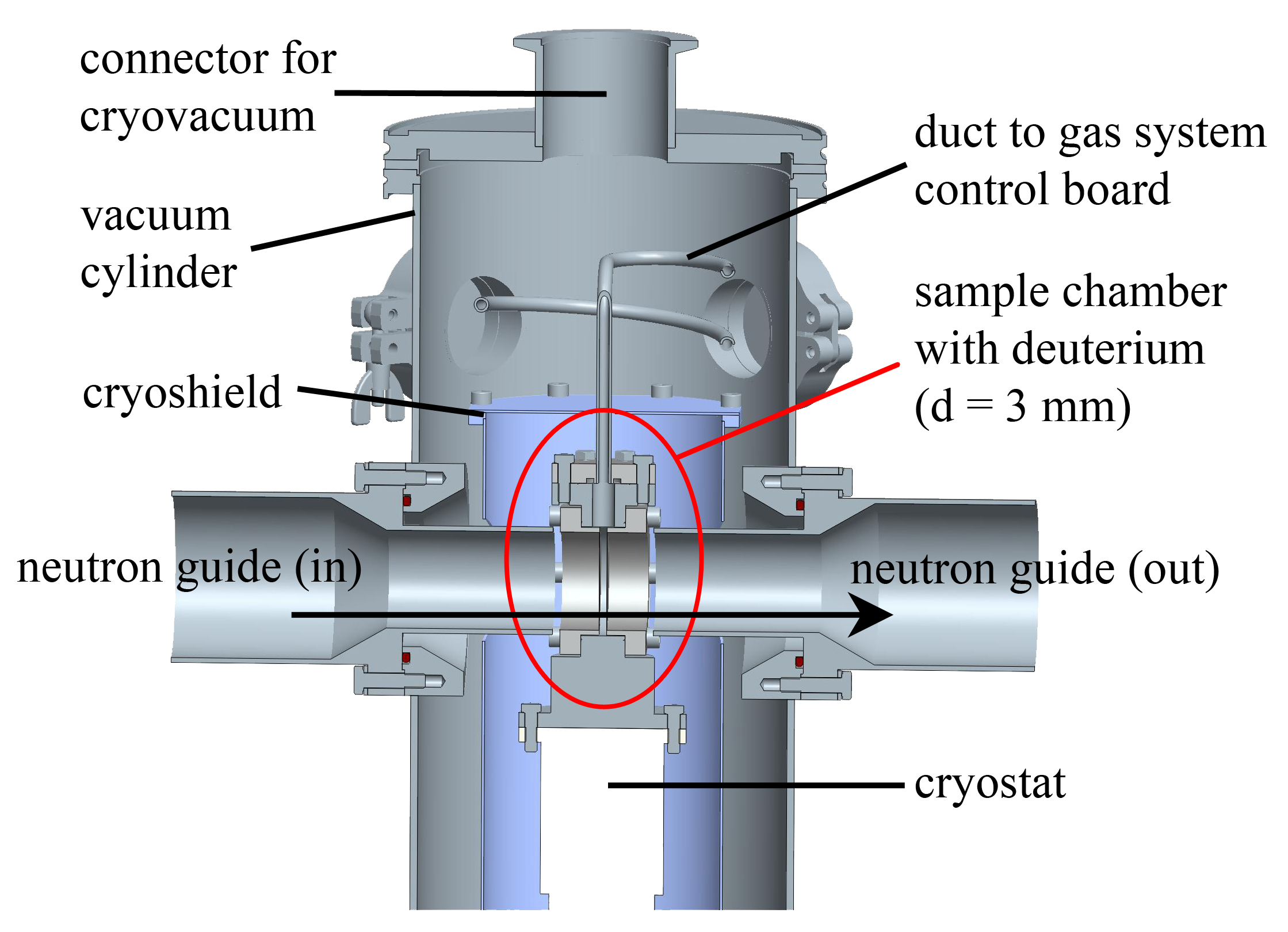}
  \caption[Sample chamber with cryo-environment and neutron guides]{Sample chamber with cryo-environment and neutron guides.\label{fig:cross-d2-gesamt}}
\end{figure}

\begin{figure}[!h]
	\includegraphics[width=1.00\columnwidth]{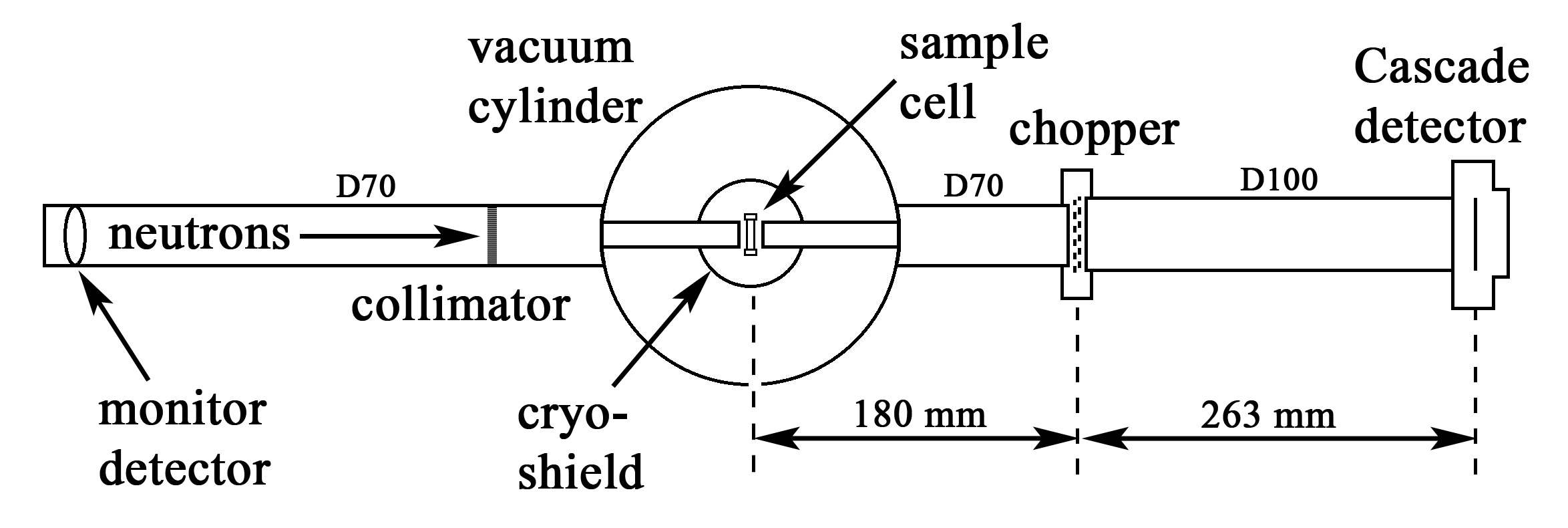}
  \caption[TOF geometry]{TOF geometry of this experiment at the PF2-EDM beamline.\label{fig:tof-geometry}}
\end{figure}

Our sample environment, see Fig.\,\ref{fig:cross-d2-gesamt}, was set up for a time-of-flight (TOF) experiment at the PF2-EDM beamline at ILL Grenoble, see Fig.\,\ref{fig:tof-geometry}. The sample cell consisted of an aluminum alloy (AlMg3) torus closed by two windows machined from aluminum (AlMg3, 0.3\,mm thick). The windows faced the UCN beam direction. This resulted in a disk-shaped liquid deuterium sample of 40\,mm diameter and 3\,mm thickness. Indium wire was used as a cryogenic sealant. Placed inside a vacuum cylinder, the sample cell was cooled by a Sumitomo closed-cycle liquid-helium cryostat (1.5\,W at 4\,K) from below and the deuterium was inserted through a feed line on its top. A resistive heater in the base of the sample cell was used for sample temperature control. The two thin-film resistance temperature sensors were located inside the torus immediately above and below the sample volume, respectively. Due to the design of the sample cell, the temperature gradient across the cell was $\Delta T = \pm 1.6\,\text{K}$ for liquid deuterium. The minimum possible sample temperature was $\sim$\,13\,K.

The incoming UCN beam was collimated by a polymethyl methacrylate (PMMA) disk (70\,mm diameter, 6\,mm thickness) with 829 holes of 1\,mm diameter each. After the collimator, the UCNs passed the sample cell and then the chopper. The latter was custom designed and consisted of two movable polyethylene grids with a trapezoidal opening function of 7\,ms FWHM and was operated in vacuum at 8\,Hz (about 1/16 duty cycle). A trigger actuated the chopper and synchronized it with the detector. The total flight path inside a tube coated with polyethylene on the inside was 263\,mm long. A CASCADE detector \citep{klein:2011} (based on $^{10}$B) with around 90\,\% detection efficiency for UCNs was mounted at the end of the experimental setup. The detector was continuously flushed with an Ar-CO$_{\text{2}}$ (90:10) gas mixture under ambient pressure and an acceleration voltage of 1200\,V was applied.

\section{Theoretical Framework}
Up until now, theoretical models of coherent and incoherent scattering of slow neutrons in deuterium have been based on semi-empirical models of the scattering law $S(q,\omega)$; for an example see Bernnat \textit{et al.} \citep{bernnat:2004}. Here we interpret these cross sections using scattering laws purely based on the properties of liquid deuterium for both the \textit{ortho} and \textit{para} spin configurations.

Theoretical calculations of the scattering cross sections for slow neutrons in \emph{gaseous} deuterium have been provided by Hamermesh and Schwinger~\citep{hamermesh:1946} and Young and Koppel~\citep{young-koppel:1964}.

Neutron scattering can be coherent or incoherent. The two can be separately described and contribute to the cross section.
\begin{equation}\label{eq:d-diff_coh-inc}
\frac{\text{d}^2 \sigma}{\text{d}\Omega \text{d}E} = \frac{1}{4\pi}\frac{k_\text{f}}{k_\text{i}} \left[\sigma^\text{coh} \times S_\text{coh}(q,\omega) + \sigma^\text{inc} \times S_\text{inc}(q,\omega) \right]
\end{equation}

\begin{figure}[!h]
	\includegraphics[width=1.00\columnwidth]{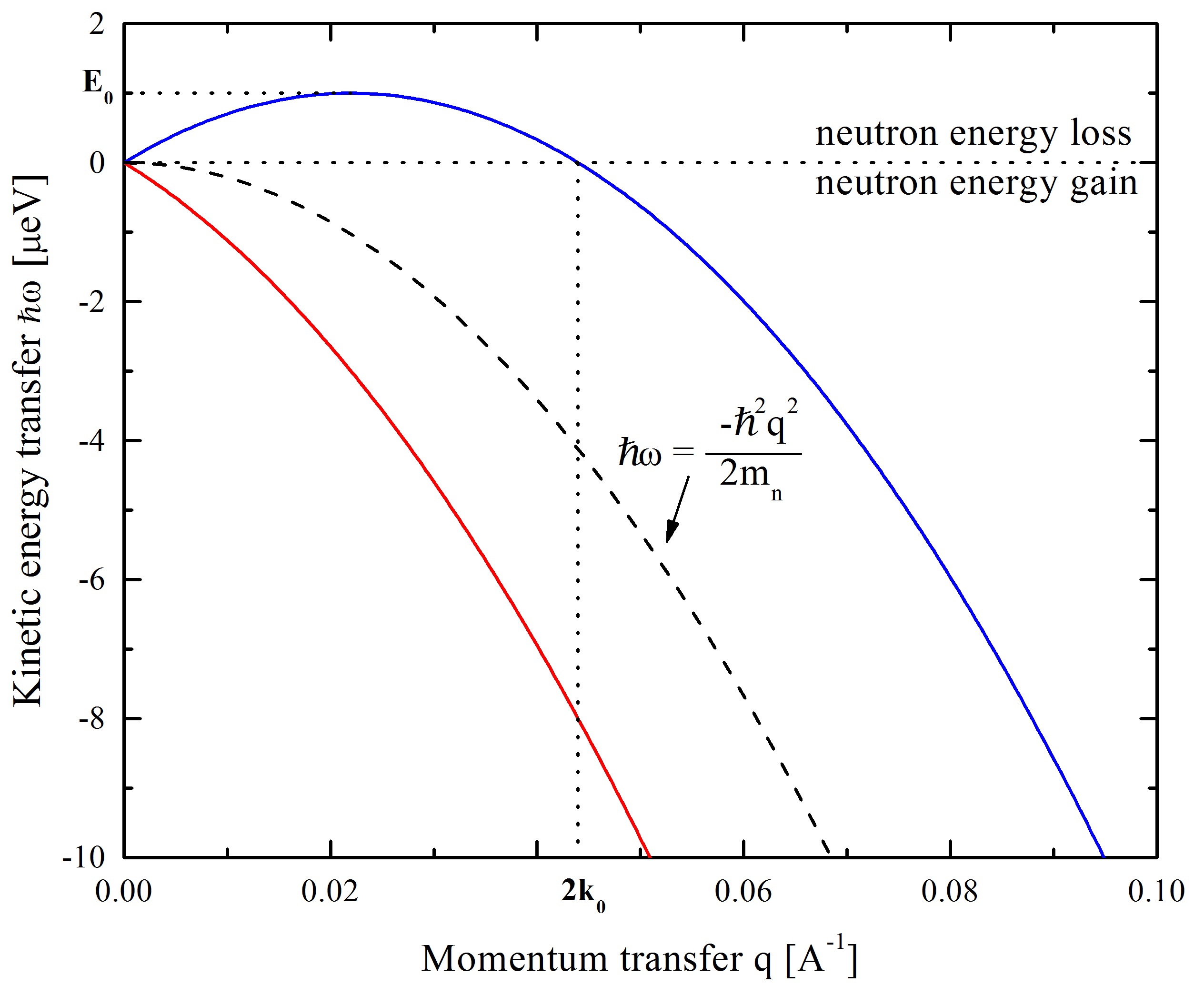}
  \caption{Graphic representation of the kinematic region (between the red (lower) and blue (upper) parabolas) for incident neutron energy $E_0=1\,\mu\text{eV}$.\label{fig:kin_region}}
\end{figure}

These double-differential cross sections have to be integrated over the kinematic region in order to obtain effective scattering cross sections:
\begin{align}
\label{eq:s-eff-formula}
\sigma_\text{eff} &= \int \left( \frac{\text{d}^2 \sigma}{\text{d}\Omega \text{d}E} \right) \text{d}\Omega \text{d}E \nonumber\\
&= \int_{-\infty}^{E_0} \int_{q_1 (E)}^{q_2 (E)} \left( \frac{1}{k_\text{i}k_\text{f}} \frac{\text{d}^2 \sigma}{\text{d}\Omega \text{d}E} \right) 2\pi q \text{d}q \text{d}E,
\end{align}
where $q_{1,2}(E)$ are the two parabolas marking off the kinematic region in the $q-E$ plane, see Fig.~\ref{fig:kin_region}:
\begin{align}
q_1(E) &= k_0\left[1 - \sqrt{1-\frac{E}{E_0}} \right]\\
q_2(E) &= k_0\left[1 + \sqrt{1-\frac{E}{E_0}} \right].
\end{align}

A measure to assess the quantum behavior of light liquids is the dimensionless de Boer quantum delocalization length \citep{deboer:1948} $\Lambda$,
\begin{equation}
\Lambda \approx \frac{\lambda_\text{de Broglie}}{a_\text{nn}} = \frac{h}{a_\text{nn}\times (2\pi m_\text{D$_2$} k_\text{B}T)^{1/2}},
\end{equation}
where $a$ is the mean nearest-neighbor separation ($a_\text{nn} = {N_\text{V}}^{-1/3} \approx \sigma_\text{HS}$), $\sigma_\text{HS}$ is the molecular hard-sphere diameter of the deuterium molecule \citep{riceboondavis:1968} (2.95\,\AA), and $\lambda_\text{de Broglie}$ is the thermal de Broglie wavelength of the deuterium molecule. For a classical particle $\Lambda \ll 1$, that is to say, the thermal de Broglie wavelength is small compared to the geometrical diameter of the particle~\citep{hansen-mcdonald:1986}. For deuterium $\Lambda=0.66$. So, deuterium can generally be regarded as a classical liquid with some quantum effects. For rotational excitations in molecules there is an additional measure to assess quantum behavior: $\Theta_\text{rot} \ll T$, where $\Theta_\text{rot} = \hbar^2/2Ik_\text{B}$ with $I$ being the molecular moment of inertia. Classical behavior is assumed when $\Theta_{\text{rot}}\ll T$ holds true. For deuterium $\Theta_\text{rot} = 22.0\,\text{K}$, hence $\Theta_\text{rot} \approx T = 20\,\text{K}$. This points to a largely quantum-mechanical behavior of the rotator states of liquid deuterium. Therefore, we will consider deuterium to be a classical liquid for the translational degrees of freedom, i.e., diffusion and phonon excitations, but with a quantum-mechanical description of the rotational states.

\subsection{Coherent scattering law}

Coherent scattering $\tilde{S}_{\text{coh}}(q,\omega)$ in our model is based on the hydrodynamic limit (HDL)~\citep{hansen-mcdonald:1986, landaulifshitz:1987}, with the speed of sound showing the known $q$-dependent dispersion in liquids, see Eq.~\ref{speed-sound}.  

A detailed treatment of the double-differential scattering cross sections is presented in Appendix~\ref{sec:appA} and the scattering laws are described in Appendix~\ref{sec:appB}.

\subsection{Incoherent scattering law}

For the incoherent scattering law $\tilde{S}_{\text{inc}}(q,\omega)$ we chose the incoherent Lovesey model~\citep{lovesey:1973-inc}. In the quasi-elastic region the incoherent Lovesey model is in good agreement with a simple Lorentzian (which represents only self-diffusion), while it properly describes the incoherent phonon contribution in the meV range. Phonons are important in our case, as the integration over the kinematic region, see Eq.~\ref{eq:s-eff-formula}, extends up to $\hbar\Omega_\text{E}$ ($\Omega_\text{E}$ is the Einstein frequency of the liquid), which is of the order of 10\,meV in liquid deuterium.

\subsection{Rotational states}

The deuterium's rotational state is described as a nearly free quantum-mechanical rotator by a Gaussian shaped scattering law $\tilde{S}^\text{rot}$ with a half width at half maximum (HWHM) of 0.45 meV for the $J$=1$\rightarrow$0 relaxation~\citep{egelstaff:1967}. Since the rotational relaxation $J$=1$\rightarrow$0 upon impact of a neutron is an incoherent scattering process \citep{liu:2000}, we again use the incoherent Lovesey model for the translational part of the scattering law. The two scattering laws need to be convolved in order to yield the correct scattering law $\tilde{S}_\text{rot}^\text{inc} = \tilde{S}^\text{inc}_\text{Lovesey} \otimes \tilde{S}^\text{rot}$ in the liquid state. We approximate the Gaussian as a delta function, because it has a significantly smaller HWHM than the quasi-elastic diffusion peak. The convolution of this peak and a delta function, however, results in that very same diffusion peak. So, as a good approximation we can simply use the incoherent Lovesey model to calculate the $J$=1$\rightarrow$0 scattering cross section. The energy $E$ will be shifted by the rotator energy \citep{souers:1986} $\hbar\omega_\text{10}$ of -7.4\,meV:
\begin{align}
\tilde{S}^\text{inc}_\text{rot} &= \tilde{S}^\text{inc}_\text{Lovesey}(q,\omega) \otimes \tilde{S}^\text{rot}(\omega) \\
&= \tilde{S}^\text{inc}_\text{Lovesey}(q,\omega) \otimes \delta (\omega - \omega_\text{10}) \\
&= \tilde{S}^\text{inc}_\text{Lovesey}(q,\omega-\omega_\text{10})
\end{align}

Most research uses the simplifying approach of the \emph{incoherent approximation} \citep{vineyard:1958, turchin:1965}, replacing the coherent scattering law $S_\text{coh}(q,\omega)$ in Eq.~\ref{eq:d-diff_coh-inc} with the incoherent one and using $\sigma_\text{tot} = \sigma^\text{coh} + \sigma^\text{inc}$. In the liquid state, $\sigma^\text{coh}$ has to be weighted with the structure factor $S^\text{HS}(q)$ to include the microscopic structure of the liquid,
\begin{equation}\label{eq:incoh-approx}
\frac{\text{d}^2 \sigma}{\text{d}\Omega \text{d}E} \simeq \frac{1}{4\pi}\frac{k_\text{f}}{k_\text{i}} \left[ (\sigma^\text{coh} \times S^\text{HS}(q) + \sigma^\text{inc}) \times S_\text{inc}(q,\omega) \right].
\end{equation}

The structure factor $S^\text{HS}(q)$ in this case is that of the Percus--Yevick approximation~\citep{boon-yip:1980}. It describes the center of mass structure factor of liquid deuterium using the hard-sphere diameter $\sigma_\text{HS}$ and agrees well with experimental data \citep{zoppi:1995} in the low-$q$ region.

The total cross section of UCNs in liquid deuterium measured in our experiment is composed of the scattering contributions from \textit{ortho}- and \textit{para}-deuterium, as well as absorption. The latter has to be corrected for in order to compare the experimental data with theory:
\begin{equation}
\sigma_\text{tot}^\text{D$_2$} = c_\text{o}\sigma_\text{00} + c_\text{p} \left(\sigma_\text{11} + \sigma_\text{10}\right) + \left( \sigma_\text{abs}^\text{D$_2$} + \sigma_\text{abs}^\text{H$_2$} \right),
\end{equation}
where $c_\text{o}$ and $c_\text{p}$ are the \textit{ortho}- and \textit{para}-deuterium concentrations. Here, $\sigma_{JJ'}$ denotes the initial ($J$) and final ($J'$) rotational state of the molecule in the scattering process.

\section{Data Treatment}
The Cascade detector control program provided files containing time-resolved (0.4\,ms per time channel) and spatially resolved (8$\times$8 pixels of 1\,cm$^2$ each) counts, which were summed up over all pixels to obtain the final TOF spectrum, normalized to one measurement run of 120\,s. The UCN time-of-flight spectra were then fitted using a constant background and two Gaussians (double Gaussian). The $R^2$ value of the relevant fits was between 0.96 and 0.98. The constant background extracted from this fit was subsequently subtracted from each measured TOF spectrum (of both empty and filled sample cells), which was then used for further calculations and data correction. About 80\,\% of the entire neutron count proved to be background, caused entirely by up-scattered (thermal) neutrons emanating from the PMMA collimator at room temperature. The average UCN count over the entire UCN spectrum was 650 per measurement run of 120\,s. The TOF spectrum was subdivided into bins of 16 time channels each. The mean time channel number was then converted into neutron velocity and, subsequently, into neutron energy.

A deconvolution of the TOF spectrum to remove the influence of the chopper was not required, because the chopper opening time of 7\,ms (FWHM) is very close to the time bin width of 6.4\,ms.

Before proceeding with the calculation of the scattering cross sections, the raw data spectrum had to be corrected for reflection and transmission on interfaces as well as for multiple scattering. We believe these two corrections are indispensable for obtaining physically correct cross sections in the UCN regime.

\subsection{Quantum-mechanical correction}

The quantum-mechanical (QM) correction takes into account reflection losses at the interface between two different potentials and provides the corrected transmission through the sample cell~\citep{steyerl:1972}. The potential of the substance through which the neutrons travel, has a real and an imaginary part. The latter accounts for absorption and inelastic scattering of UCNs and is so small that it can be considered negligible. Transmission through a potential interface and the reflection from it are described by
\begin{equation}
T=\frac{4k\text{Re}(k')}{|k+k'|^2}%\text{ and}
\end{equation}
and
\begin{equation}
R=\frac{|k-k'|^2}{|k+k'|^2},
\end{equation}
where $k$ is the wave vector of the incident neutron outside the potential wall (i.e., the deuterium sample), and $k'$ is the neutron wave vector inside the potential wall.

The total reflection $\rho$ and transmission $\tau$ through the entire potential wall can be calculated from
\begin{equation}
\rho=R\frac{1+\alpha^2 (T-R)}{1-\alpha^2 R^2}%\text{ and}
\end{equation}
and
\begin{equation}
\tau=\frac{\alpha T^2}{1-\alpha^2 R^2},
\end{equation}
where $\alpha = \text{exp}(-\sigma_\text{tot} N_\text{V} d)$ is the attenuation coefficient. It describes the decreased neutron count depending on the total cross section and the thickness of the sample.

The experimental transmission is then corrected as follows:
\begin{equation}
\tau^\text{expt} = \left( \frac{I_\text{filled}}{I_\text{vacuum}} \right)^\text{expt} = \frac{\alpha^\text{QM corr} T^2}{1-\left( \alpha^\text{QM corr}\right)^2 R^2}.
\end{equation}
This equation is numerically solved for $\alpha^\text{QM corr}$, which represents the quantum-mechanically \emph{corrected} experimental transmission,
\begin{equation}\label{qm-corr-alpha}
\alpha^\text{QM corr} = \left( \frac{I_\text{filled}}{I_\text{vacuum}} \right)^\text{QM corr},
\end{equation}
i.e., as if no reflection losses had occurred at the potential interface.

\subsection{Multiple scattering correction}

\begin{figure}[!h]
	\includegraphics[width=1.00\columnwidth]{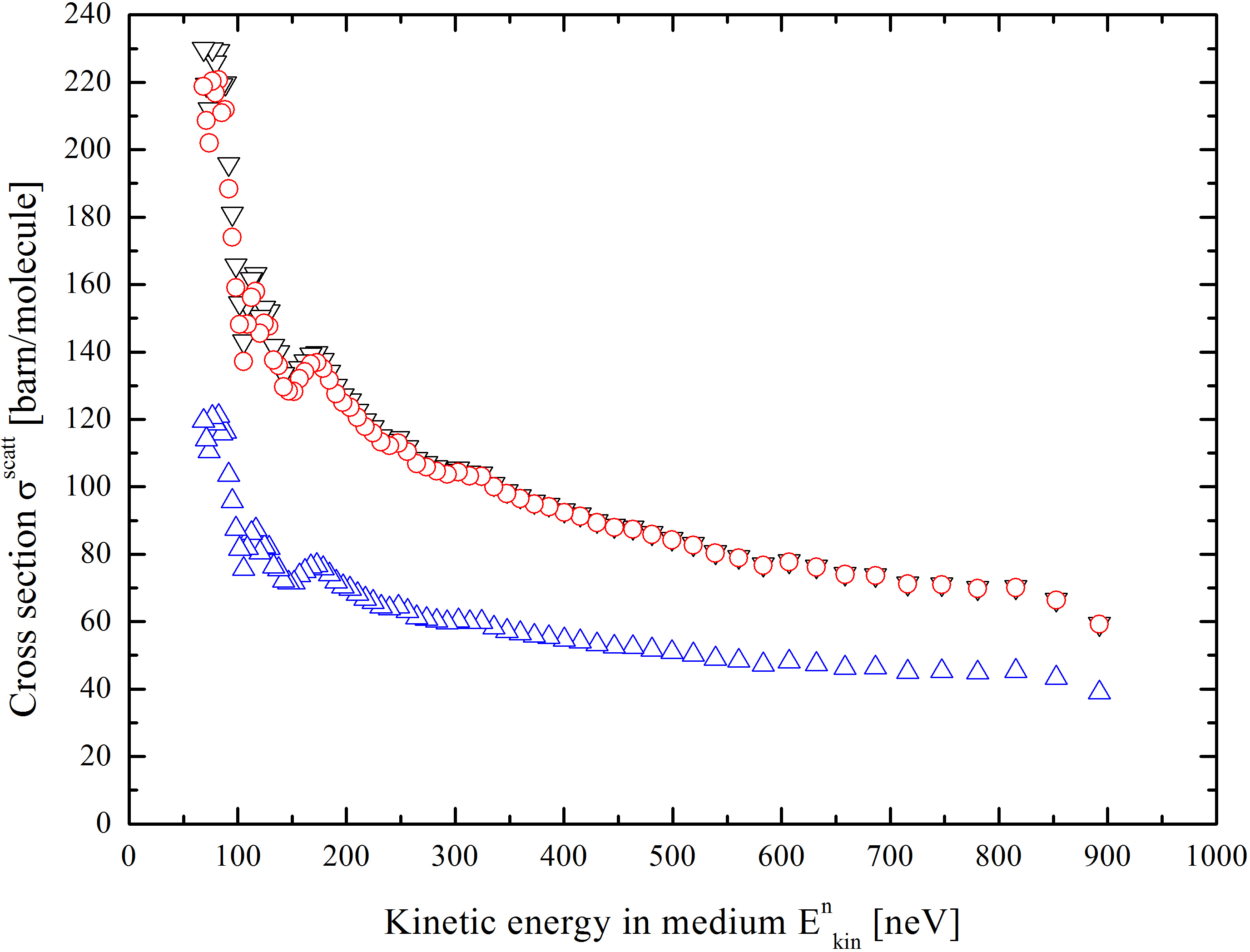}
  \caption{Comparison of scattering cross sections without correction (black down triangles $\bigtriangledown$), with QM correction (red circles {\color{red}$\bigcirc$}) and with corrections for both QM effects and multiple scattering (blue triangles {\color{blue}$\triangle$}) for $T = 20.6\,\text{K}$.\label{fig:Vgl_20-6K_mitohneKorr_konstSears_KorrVgl}}
\end{figure}

The corrected transmission $\alpha^\text{QM corr}$ is then used in the multiple scattering (MS) correction. From Fig.~\ref{fig:Vgl_20-6K_mitohneKorr_konstSears_KorrVgl} it becomes clear that the quantum mechanical correction has an effect only for very slow neutrons with energies below 200\,neV, which is close to the optical potential (also called the Fermi potential) of liquid deuterium at 20.6\,K, $V_\text{F}=88.5\,\text{neV}$ (obtained from theoretical calculations \citep{fermi:1934}). For reasons of simplicity the QM correction was carried out only for the potential barrier vacuum--deuterium--vacuum, neglecting the aluminum sample cell windows. After a more thorough calculation for the vacuum--aluminum--deuterium--aluminum--vacuum potential \citep{herold:2013}, it was found that neglect of the aluminum barrier results in a final cross-section error of only 1\,\%-2\,\%.

Multiple scattering cannot be avoided, because the criterion $\Sigma d < 1$ can hardly be met in a UCN transmission experiment in liquid deuterium with $\Sigma$ of the order of 3\,cm$^{-1}$. Thus, even in very thin samples, such as ours, multiple scattering occurs and has to be corrected for. In this step we use the count ratio corrected for QM effects (superscript \emph{QMcorr}) from Eq.~\ref{qm-corr-alpha} and calculate the count ratio corrected for multiple scattering (superscript \textit{1}). Sears' theory of multiple scattering~\citep{sears:1975}, which we think provides the best tool for this case, yields the \emph{single}-scattering cross section $\sigma^1$:
\begin{align}
\left( \frac{I_{\text{filled}}}{I_{\text{vacuum}}} \right)^1 &= e^{-\sigma^1 N_\text{V} d}\\
&= \frac{\left[ \left( \frac{I_{\text{filled}}}{I_{\text{vacuum}}} \right)^{\text{QMcorr}} \right] + \Delta(\Sigma d)}{1+\Delta(\Sigma d)},\label{eq:mult-corr}
\end{align}
where $\Delta$ denotes the ratio of multiple ($n\geq2$) to single scattering,
\begin{equation}
\Delta = \frac{I_{\text{n$\geq$2}}}{I_{\text{n=1}}} = \frac{e^{2\delta} -1}{2\delta}-1,
\end{equation}
with
\begin{equation}\label{eq:delta-of-Sigma-d}
\delta = \frac{1}{2} \Sigma d \left[ c^{*} - \ln (\Sigma d) + \frac{1}{3}\Sigma d + O\left( (\Sigma d)^2 \times \ln (\Sigma d) \right)\right]
\end{equation}
and
\begin{equation}
c^{*} = 0.92278,\, \Sigma = \sigma^1 N_{\text{V}}
\end{equation}

Equation~\ref{eq:mult-corr} needs to be solved numerically for $\sigma^1$. As $\delta (\Sigma d)$, see Eq.~\ref{eq:delta-of-Sigma-d}, ceases to yield reliable results for $\Sigma d \approx 1$~\citep{sears:1975}, we used this formula only up to $\Sigma d = 0.7$, i.e., for energies down to $E_\text{kin} = 430\,\text{neV}$. We then linearly extrapolated the smoothly decreasing correction factor (which equals the MS-corrected ({\color{blue}$\triangle$}) scattering cross section over the QM-corrected ({\color{red}$\bigcirc$}) scattering cross section, see Fig.~\ref{fig:Vgl_20-6K_mitohneKorr_konstSears_KorrVgl}) down to $E_\text{kin} = 70\,\text{neV}$ to obtain the final single scattering cross section $\sigma^1$.

Hydrogen impurities with a concentration of $c_\text{H$_\text{2}$}=0.0025$ in the deuterium account for some losses in the liquid. They contribute an absorption cross section of $c_\text{H$_\text{2}$} \times (730\,\text{b}\times \text{m/s})/v$. At $T=20\,K$, close to 100\,\% of the hydrogen are in the $J$=0 para-state, which scatters only coherently with $\sigma^{\text{p-H}_2}(v)=c_{\text{H}_2} \times (191\,\text{b}\times \text{m/s})/v$. Absorption in deuterium equals $(1.1\,\text{b}\times \text{m/s})/v$. These contributions are negligible in comparison with the scattering cross sections.

\section{Results and experimental cross sections}
Neutron kinetic energies are given as in-medium energies. Therefore, the Fermi potential of liquid deuterium was subtracted from the kinetic energy measured with the TOF geometry $E^{\text{in medium}}=E_{\text{kin}}-V_{\text{F}}$.

\begin{figure}[!h]
	\includegraphics[width=1.00\columnwidth]{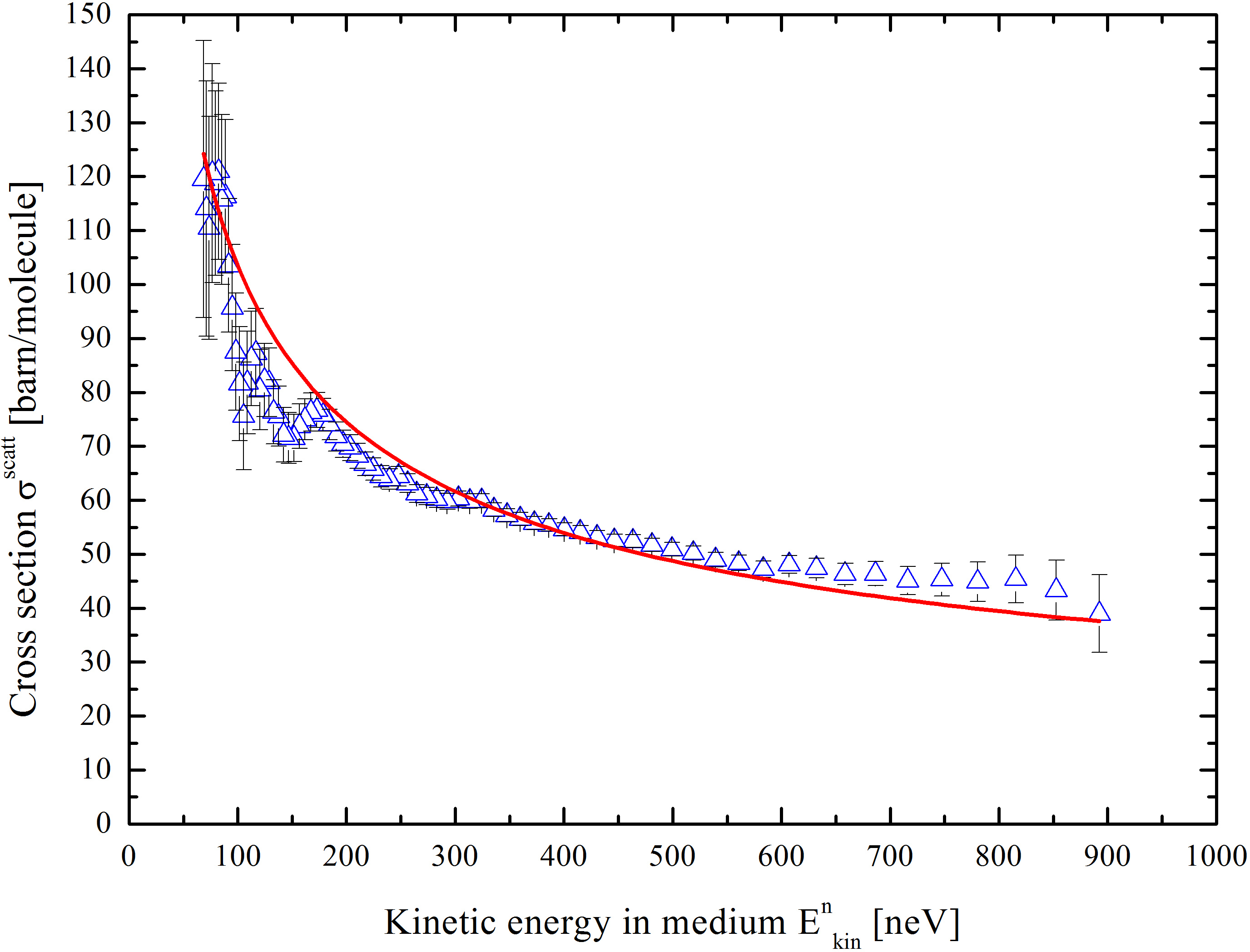}
  \caption{Fully corrected experimental data (blue triangles {\color{blue}$\triangle$}), and theoretical model for an 80\,\% ortho- and 20\,\% para-deuterium mixture (solid red line {\color{red}---}) for $T = 20.6\,\text{K}$.\label{fig:Vgl_20-6K_mitKorr_konstSears_Modell}}
\end{figure}

Figure~\ref{fig:Vgl_20-6K_mitKorr_konstSears_Modell} shows the corrected experimental data and the calculated model for the 80\,\% ortho- and 20\,\% para-deuterium mixture from our experiment at 20.6\,K. It is obvious that our model and experimental results overlap almost completely.

\begin{figure}[!h]
	\includegraphics[width=1.00\columnwidth]{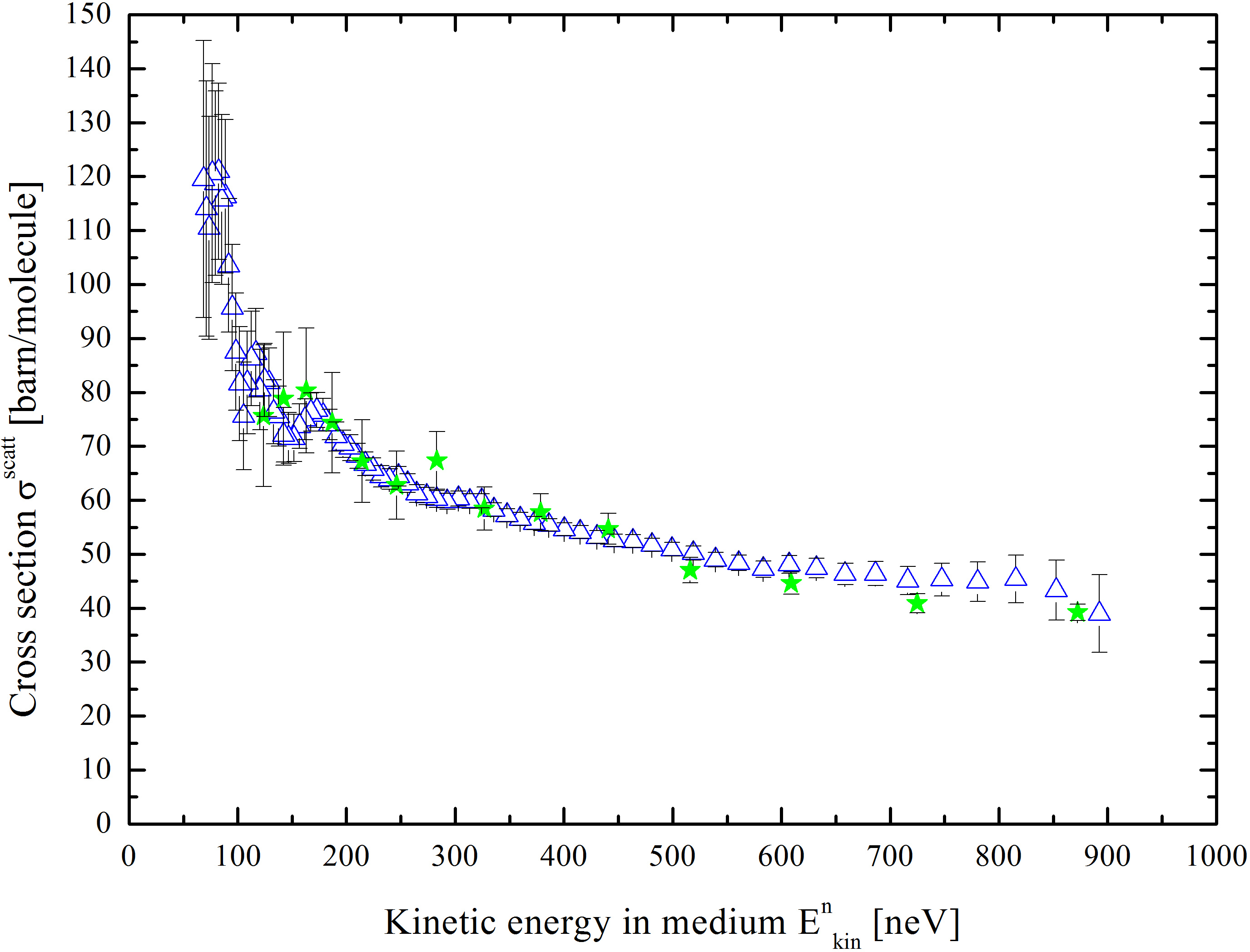}
  \caption{Scattering cross sections of liquid deuterium from the Atchison \textit{et al.} experiment at $T = 19\,\text{K}$ (green stars {\color{green}$\bigstar$}) as well as from this experiment at $T = 20.6\,\text{K}$ (blue triangles {\color{blue}$\triangle$}).\label{fig:schweizer-ourexp}}
\end{figure}

As can be seen in Fig.~\ref{fig:schweizer-ourexp}, the scattering cross sections reported here for liquid D$_{\text{2}}$ at 20.6\,K overlap well with the Atchison \textit{et al.} data \citep{atchison:2005-liq} for 19\,K. The temperature uncertainty reported by Atchison \textit{et al.}, (19.0$\pm$0.1)\,K, refers to the deviation from the mean cell temperature. Their temperature gradient across the cell \citep{kasprzak:2014} was 0.5\,K.
%The paper by Bodek \citep{bodek:2004}, wherein the PSI transmission instrument is described in detail, mentions four temperature sensors, but neither their position on the sample cell nor the minimum achievable temperature gradient across the cell. So it remains unclear, whether the reported (19$\pm$0.1)\,K may actually have a larger uncertainty. Two more experiments using the same equipment report three \citep{atchison:2005-thermo} and one temperature sensor on the sample cell \citep{kasprzak:2008}, respectively.
When calculated for 19\,K, our model delivers cross sections of about 10\,\%-15\,\% below the data of Atchison \textit{et al.}, with increasing agreement for higher neutron energies.

It is worth noting that both data sets independently show a peak around $E_\text{kin} = 170\,\text{neV}$, which has not been noticed previously. It stems very probably from the hyperfine splitting (hfs) of the ground state of the deuterium molecule. While there are apparently no published experimental hfs data for deuterium, data are available for the hydrogen molecule, which exhibits a low-lying hfs of 1008\,neV (244\,MHz) \citep{frey:1962}. Scaling this value with the ratio $\text{g}_\text{D}/\text{g}_\text{P} = 0.15$ -- where $\text{g}_\text{D,P}$ are the molecular $g$-factors of the deuteron and the proton, respectively -- we arrive at an estimated hfs of 150\,neV for the deuterium molecule. Hence, the peak in question, see Figs.~\ref{fig:Vgl_20-6K_mitKorr_konstSears_Modell} and \ref{fig:schweizer-ourexp}, may represent the lowest-lying internal inelasticity of the deuterium molecule, which is due to hfs of its ground state. A possible quadrupole shift is negligible in deuterium molecules due to the marginal quadrupole interaction energy of 0.93\,neV \citep{kolsky:1952}.

\begin{figure}[!h]
	\includegraphics[width=1.00\columnwidth]{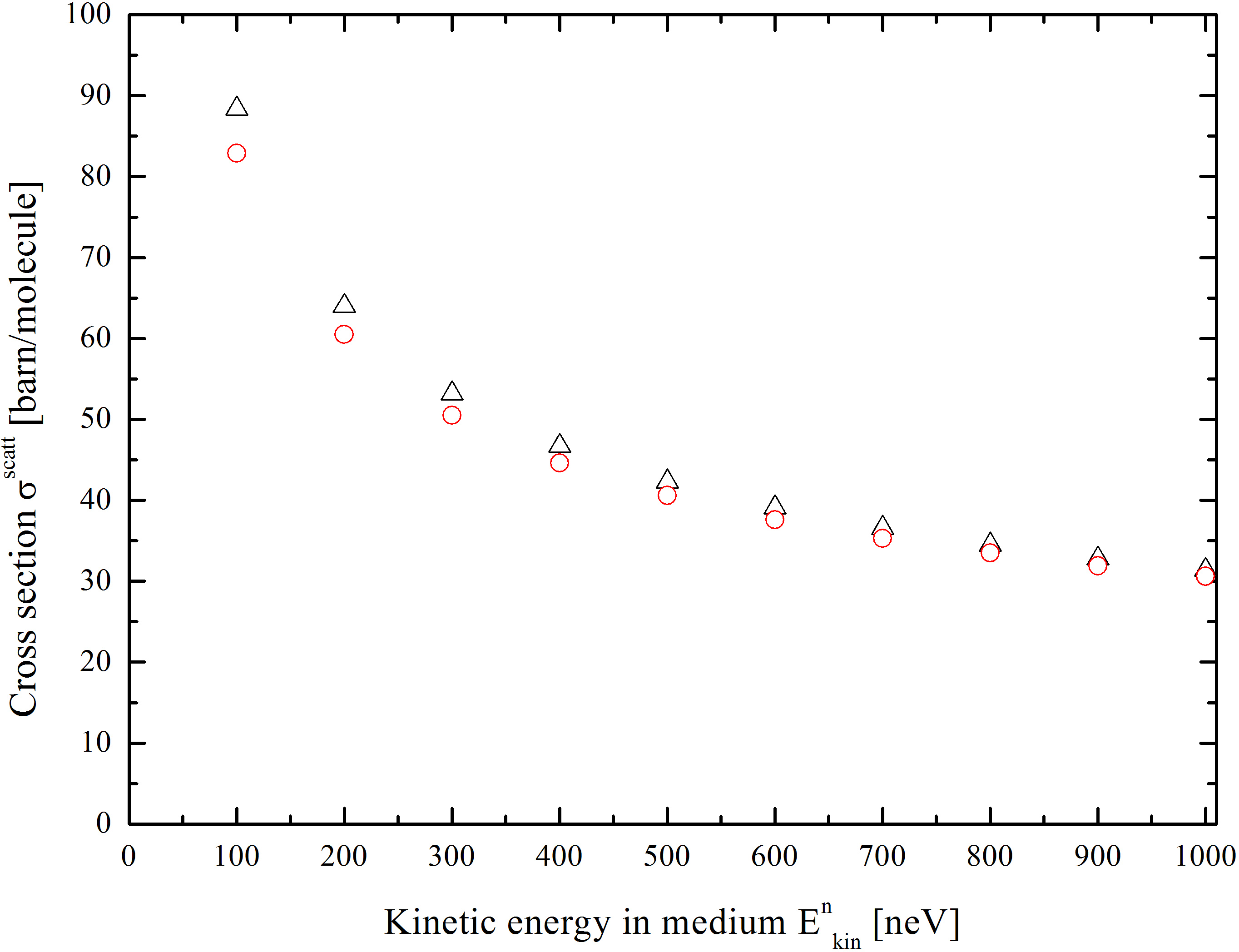}
  \caption{Liquid deuterium ($c_{\text{o}}=1.0$) scattering cross sections for $T=19\,\text{K}$ using our exact model (black triangles $\triangle$) and the incoherent approximation (red circles {\color{red}$\bigcirc$}).\label{fig:Vgl_exakt-incoh-approx}}
\end{figure}

The comparison between the exact calculation of the coherent and incoherent scattering cross sections for liquid \textit{ortho}-deuterium on the one hand and the \emph{incoherent approximation} (see Eq.~\ref{eq:incoh-approx}) on the other shows that the latter does provide acceptable results. Its scattering cross sections lie 7\,\% below those of the exact model for $E_\text{kin} = 100\,\text{neV}$ and become even more accurate with higher neutron energies, as is shown in Fig.~\ref{fig:Vgl_exakt-incoh-approx}.

\section{Conclusions}
The models for coherent and incoherent scattering in liquid deuterium developed in this work possess the correct 1/$v$ behavior in the UCN limit. They are in excellent agreement with our experimental data and in acceptable agreement with the experimental data from Atchison \textit{et al.} for the respective temperatures. It is worth noting that the models do not contain any free scaling parameters, but yield the scattering cross sections on an absolute scale. The experimental data were properly evaluated and the relevant corrections for quantum-mechanical transmission and multiple scattering were applied.

When comparing the exact models with the commonly used incoherent approximation, it was found that the latter holds true for liquid deuterium at low temperatures. The incoherent approximation proved to be quite accurate with a scattering cross section only 7\,\% below our exact model for $E_\text{kin} = 100\,\text{neV}$ and an even better agreement for higher neutron energies.

Future investigations into this matter will first have to verify the deuterium self-diffusion coefficient $D_\text{s}$, since the literature value has to be used with caution. A reliably measured self-diffusion coefficient would serve as a means to validate our incoherent model.

Due to the surface roughness of the sample cell's aluminum windows, some boundary effects may have arisen from the interaction of deuterium with the surface texture or with the crystal structure of the aluminum windows. This might have reduced UCN transmission.

The models provided here are viable tools for the calculation of liquid deuterium scattering cross sections for arbitrary \textit{ortho}- and \textit{para}-concentrations at low temperatures from UCN energies up to the meV range. They are easily applicable to liquid hydrogen as well.

\appendix
\section{Cross sections\label{sec:appA}}
According to theory \citep{hamermesh:1946, young-koppel:1964, liu:2000}, the bound \emph{nuclear} scattering cross sections per deuterium molecule, which will later be used in conjunction with scattering laws to calculate the total scattering cross section for deuterium in the liquid state, are
\begin{subequations}
\begin{align}
\sigma_{00}^\text{inc} &= 4\pi \times 2 \left(\frac{5}{4}b_\text{inc}^2\right) = 5.12\,\text{barn}\label{eq:inc-00-sigma},\\
\sigma_{00}^\text{coh} &= 4\pi \times 4 (b_\text{coh}^2) = 22.4\,\text{barn},\\
\sigma_{11}^\text{inc} &= 4\pi \times 2 \left(\frac{1}{2}b_\text{inc}^2\right) = 2.05\,\text{barn}\label{eq:inc-11-sigma},\\
\sigma_{11}^\text{coh} &= 4\pi \times 4 (b_\text{coh}^2) = 22.4\,\text{barn},\\
\sigma_{10}^\text{inc} &= 4\pi \times 2 \left(\frac{3}{2}b_\text{inc}^2\right) = 6.15\,\text{barn},
\end{align}
\end{subequations}
with $b_\text{coh} = 0.6671\times 10^{-12}\,\text{cm}$ and $b_\text{inc} = 0.404\times 10^{-12}\,\text{cm}$.

The values for the bound scattering lengths $b_\text{coh}$ and $b_\text{inc}$ were taken from V. F. Sears \citep{sears:1992}. Note that, for instance, the molecular $\sigma_{00}^\text{inc}$ is not simply the incoherent scattering cross section of one deuteron multiplied by two. It is larger than that, due to the interference of incoherently scattered neutron waves from both spin-correlated atoms within the same molecule.

Deuterium is a homonuclear diatomic molecule and looks like a dumbbell. Squared Bessel functions describe the symmetry of the deuterium molecule's wave function for different rotational states ($J$,$J'$) and act as form factors of the molecule. They account for the equilibrium separation of the deuterons in the molecule $a = 0.74\,\text{\AA}$ and have to be multiplied by the nuclear scattering cross sections, see Eqs.~\ref{eq:s00} to \ref{eq:s10}, depending on the rotational transition of the deuterium molecule:
\begin{subequations}
\begin{align}
&J\text{=0$\rightarrow$0:}\; \sigma_{00} \times \left[j_0^2\left(\frac{aq}{2}\right)\right], \\
&J\text{=1$\rightarrow$1:}\; \sigma_{11} \times \left[j_0^2\left(\frac{aq}{2}\right) + 2 j_2^2\left(\frac{aq}{2}\right)\right], \\
&J\text{=1$\rightarrow$0:}\; \sigma_{10} \times \left[j_1^2\left(\frac{aq}{2}\right)\right].
\end{align}
\end{subequations}

\begin{widetext}

For liquid deuterium the scattering cross sections are quasielastic (\emph{qel}) and inelastic (\emph{\textpm 1ph, 0ph/{-1}rot}). The 0ph/\mbox{{-1}rot} process (rotational relaxation without phonon creation) scatters the neutron inelastically, because it receives all of the roton energy \citep{liu:2000}. The double-differential cross sections, which are subsequently inserted into Eq.~\ref{eq:s-eff-formula}, are calculated as follows
\begin{flalign}
\label{eq:s00}
\left( \frac{\text{d}^2 \sigma}{\text{d}\Omega \text{d}E} \right)_\text{00}^\text{tot} &= \left( \frac{\text{d}^2 \sigma}{\text{d}\Omega \text{d}E} \right)_\text{00}^\text{qel,coh} + \left( \frac{\text{d}^2 \sigma}{\text{d}\Omega \text{d}E} \right)_\text{00}^\text{qel,inc} + \left( \frac{\text{d}^2 \sigma}{\text{d}\Omega \text{d}E} \right)_\text{00}^\text{{-1}ph,coh} + \left( \frac{\text{d}^2 \sigma}{\text{d}\Omega \text{d}E} \right)_\text{00}^\text{{-1}ph,inc} & \nonumber\\
&= \sqrt{\frac{E}{E_0}}\frac{1}{4\pi}e^{\frac{\hbar \omega}{2 k_\text{B}T}} \left[ \sigma^\text{coh}_{00} \times j_0^2 \left(\frac{aq}{2}\right) \times \tilde{S}^\text{coh}_\text{HDL} (q,\omega) + \sigma^\text{inc}_{00} \times j_0^2 \left(\frac{aq}{2}\right) \times \tilde{S}^\text{inc}_\text{Lovesey} (q,\omega) \vphantom{\sum} \right]
\end{flalign}

\begin{multline}
\label{eq:s11}
\hspace{-0.4cm}\left( \frac{\text{d}^2 \sigma}{\text{d}\Omega \text{d}E} \right)_\text{11}^\text{tot} = \left( \frac{\text{d}^2 \sigma}{\text{d}\Omega \text{d}E} \right)_\text{11}^\text{qel,coh} + \left( \frac{\text{d}^2 \sigma}{\text{d}\Omega \text{d}E} \right)_\text{11}^\text{qel,inc} + \left( \frac{\text{d}^2 \sigma}{\text{d}\Omega \text{d}E} \right)_\text{11}^\text{-1ph,coh} + \left( \frac{\text{d}^2 \sigma}{\text{d}\Omega \text{d}E} \right)_\text{11}^\text{-1ph,inc} \\
\left. = \sqrt{\frac{E}{E_0}}\frac{1}{4\pi}e^{\frac{\hbar \omega}{2 k_\text{B}T}} \left[ \sigma^\text{coh}_{11} \times \left\{ j_0^2\left(\frac{aq}{2}\right) + 2j_2^2 \left(\frac{aq}{2}\right)\right\} \times \tilde{S}^\text{coh}_\text{HDL} (q,\omega) + \sigma^\text{inc}_{11} \times \left\{ j_0^2\left(\frac{aq}{2}\right) + 2j_2^2 \left(\frac{aq}{2}\right)\right\} \times \tilde{S}^\text{inc}_\text{Lovesey} (q,\omega) \vphantom{\sum} \right]\right.
\end{multline}

\begin{flalign}
\label{eq:s10}
\left( \frac{\text{d}^2 \sigma}{\text{d}\Omega \text{d}E} \right)_\text{10}^\text{tot} &= \left( \frac{\text{d}^2 \sigma}{\text{d}\Omega \text{d}E} \right)_\text{10}^\text{-1ph,inc} + \left( \frac{\text{d}^2 \sigma}{\text{d}\Omega \text{d}E} \right)_\text{10}^\text{0ph/-1rot,inc} + \left( \frac{\text{d}^2 \sigma}{\text{d}\Omega \text{d}E} \right)_\text{10}^\text{+1ph,inc} & \nonumber\\
&= \sqrt{\frac{E}{E_0}}\frac{1}{4\pi}\vphantom{\sqrt{\frac{E}{E_0}}}e^{\frac{\hbar \omega}{2 k_\text{B}T}} \left[ \vphantom{\sqrt{\frac{E}{E_0}}}\sigma^\text{inc}_{10} \times \vphantom{\sqrt{\frac{E}{E_0}}}j_1^2 \left(\frac{aq}{2}\right) \times \vphantom{\sqrt{\frac{E}{E_0}}}\tilde{S}^\text{inc}_\text{Lovesey}(q,\omega-\omega_{10})\right]
\end{flalign}

\end{widetext}

In Eqs. \ref{eq:s00} to \ref{eq:s10}, the factor $e^{\frac{\hbar \omega}{2 k_\text{B}T}}$ is the \emph{asymmetry factor} \citep{sears:1985}, which connects the symmetrical model $\tilde{S}(q,\omega)$ with the physically correct scattering law $S(q,\omega)$. It suppresses the respective scattering law $\tilde{S}(q,\omega)$ on the lower half-plane of the kinematic region in Fig.~\ref{fig:kin_region}:
\begin{equation}
S(q,\omega) = \tilde{S}(q,\omega) \times e^{\frac{\hbar \omega}{2 k_\text{B}T}}
\end{equation}

In this experiment we have to deal with deuterium only in its ground state (\textit{ortho} ``00," \textit{para} ``11," and the rotational transition ``10"), because higher energy levels are not populated at temperatures as low as 20\,K. For example, the $J$=2 \textit{ortho}-state ($\omega_{20}=22.2\,\text{meV}$), which could relax into the $J$=1 \textit{para}-state, has a population of
\begin{equation}
e^{\frac{-\hbar \omega_{20}}{k_\text{B}T}} \approx e^{-13} \approx 2\times 10^{-6}
\end{equation}
\noindent
and need not to be considered. The UCNs on the other hand have too little kinetic energy to excite the deuterium molecules.

\section{Scattering laws\label{sec:appB}}
The \emph{incoherent Lovesey model} \citep{lovesey:1973-inc}, used here to describe incoherent neutron scattering in liquid deuterium, covers self-diffusion and phonons. Its scattering law is
\begin{equation}
\resizebox{.98\hsize}{!}{$\tilde{S}^\text{inc}_\text{Lovesey} (q,\omega) = \frac{1}{\pi \hbar} \frac{\tau_\text{s} \times \omega_\text{s}^2 (2\omega_\text{s}^2 + \Omega_\text{E}^2)}{\omega^2 \times \tau_\text{s}^2(q) \left[ \omega^2 - 3\omega_\text{s}^2 - \Omega_\text{E}^2\right]^2 + \left[ \omega^2 - \omega_\text{s}^2 \right]^2}$}
\end{equation}
with
\begin{equation}
%\text{with }
\omega_\text{s}^2(q) = \frac{k_\text{B}T q^2}{m_\text{D$_2$}}%\text{ and}
\end{equation}
and
\begin{equation}
\tau_\text{s}(q) = \frac{k_\text{B}T}{m_\text{D$_2$} D_\text{s} \Omega_\text{E}} \frac{1}{\sqrt{2\omega_\text{s}^2 + \Omega_\text{E}^2}}.
\end{equation}

By using the values for $D_\text{s}$ (in units of $\left[\frac{\text{cm}^2}{\text{s}}\times 10^{-5}\right]$) from Souers~\citep{souers:1986}, i.e., $D_\text{s}(19\,\text{K}) = 3.32$ and $D_\text{s}(20.6\,\text{K}) = 4.14$, unsatisfactory results were obtained. After replacement with the self-diffusion coefficients measured by E. Gutsmiedl~\citep{gutsmiedl:2013}, i.e. $D_\text{s}(19\,\text{K}) = 1.80$ and $D_\text{s}(22\,\text{K}) = 2.80$, the results for incoherent scattering agreed much better with the experimental data. $\Omega_\text{E}$ represents the Einstein frequency, which can be derived from the Debye temperature for liquid deuterium \citep{doege:2014} $\Theta_\text{D}=91\,\text{K}$ through the following relation~\citep{nachtrieb:1976}
\begin{equation}
\omega_\text{D} = \frac{k_\text{B}\Theta_\text{D}}{\hbar} = \frac{4}{3}\Omega_\text{E}.
\end{equation}

The \emph{HDL model} \citep{hansen-mcdonald:1986} is used to calculate the coherent scattering contribution. For the derivation see Landau and Lifshitz \citep{landaulifshitz:1987}. The coherent scattering law is
\begin{multline}\label{eqn:hdl-model}
\tilde{S}_\text{HDL}^\text{coh}(q,\omega)= \frac{1}{\pi}S^\text{HS}(q) \left[ \left( \frac{\gamma - 1}{\gamma}\right) \frac{D_\text{h}q^2}{\omega^2 + (D_\text{h}q^2)^2} +\right.\\
\left. \resizebox{.85\hsize}{!}{$\frac{1}{2\gamma} \left( \frac{\Gamma q^2}{(\omega + c_\text{s}(q)q)^2 + (\Gamma q^2)^2} + \frac{\Gamma q^2}{(\omega - c_\text{s}(q)q)^2 + (\Gamma q^2)^2} \right) $} \right],
\end{multline}
where $\frac{\gamma - 1}{\gamma}$ is the weighting factor for the central peak and $\frac{1}{2\gamma}$ that for each of the phonons; and $\gamma = 1.82$ according to Souers \citep{souers:1986}. $c_\text{s}(q)$ is the speed of sound, showing the following dependence on $q$:
\begin{equation}\label{speed-sound}
c_\text{s}(q) = \sqrt{\frac{\gamma k_\text{B}T}{m_\text{D$_\text{2}$} S^\text{HS}(q)}}.
\end{equation}

The accuracy of the equation above can be checked by inserting $T = 20\,\text{K}$ and $S(q$$\rightarrow$$0)=0.063$ from Zoppi \citep{zoppi:1995}. $c_\text{s}(q$$\rightarrow$$0)=1090\,\text{m/s}$ compares excellently to $c_\text{s}=1060\,\text{m/s}$ at the same temperature \citep{souers:1986}.%, p. 70.

The phonon half width at half maximum (HWHM) $\omega_\text{1,2}(q)$ is given by $\Gamma(q)q^2$. This quantity is taken from the coherent Lovesey model~\citep{lovesey:1971-coh}. Thus we plug into Eq.~\ref{eqn:hdl-model}
\begin{equation}
\Gamma(q)q^2 = \frac{1}{2\tau_\text{L}(q)} \left(1 - \frac{\omega_0^2(q)}{\omega_\text{L}^2(q)} \right),%\text{, where}
\end{equation}
where
\begin{equation}
\omega_0^2(q) = \frac{k_\text{B}T q^2}{m_\text{D$_\text{2}$} S^\text{HS}(q)}%\text{ and}
\end{equation}
and
\begin{multline}
\omega_\text{L}^2(q) = 3\omega_0^2(q) + \Omega_\text{E}^2 \left(1- 3\frac{\sin (q\sigma)}{(q\sigma)} - \right.\\
\left. 6\frac{\cos (q\sigma)}{(q\sigma)^2} + 6\frac{\sin (q\sigma)}{(q\sigma)^3} \right)
\end{multline}
and the relaxation time $\tau_\text{L}(q)$ is taken from Lovesey~\citep{lovesey:1971-coh}.
%\begin{equation}
%\tau_\text{L}(q) = \frac{\sqrt{\pi}}{2}\frac{1}{\sqrt{\omega_\text{L}^2 - \omega_0^2}}
%\end{equation}

\begin{acknowledgments}
This work was supported by Lehr\-stuhl E21 and E18 at the Physik-Department of Tech\-ni\-sche Uni\-ver\-si\-t{\"a}t M{\"u}n\-chen, the Maier-Leibniz-Laboratorium (MLL) and the For\-schungs\-neu-tro\-nen\-quelle Heinz Maier-Leibniz (FRM II). We express our gratitude to T. Deuschle, A. Frei, T. Brenner and T. Zechlau as well as to ILL and UCN-H\"{u}tte staff for their support before and during the experiment. We thank P. Hartung for the design of the cryo-environment and M. Kasprzak for providing the PSI experimental data.
\end{acknowledgments}


\end{document}